\documentclass[manuscript]{aastex}

\shorttitle{Why are the magnetic field directions measured by \emph{Voyager 1} 
on both sides of the heliopause so similar?}

\shortauthors{Grygorczuk et al.}

\begin{document}


\title{Why are the magnetic field directions measured by \emph{Voyager 1} 
on both sides of the heliopause so similar?}




\author{J. Grygorczuk}
\affil{Space Research Centre, Warsaw, Poland}
\email{jolagry@cbk.waw.pl}

\author{A. Czechowski and S. Grzedzielski}
\affil{Space Research Centre, Warsaw, Poland}




\begin{abstract}

The solar wind carves in the interstellar plasma a cavity bounded by 
a surface, called the heliopause (HP), that separates the plasma and magnetic 
field of solar origin from the interstellar ones.  It is now generally
accepted that in August 2012 
\emph{Voyager 1} (V1) crossed that boundary. 
Unexpectedly, the magnetic fields on both its sides, although theoretically 
independent of each other, were found to be similar in direction. 
This delayed the identification of the boundary as the heliopause and led to 
many alternative explanations.
Here we show that the \emph{Voyager 1} observations can be readily explained 
and, after the \emph{Interstellar Boundary Explorer} (\emph{IBEX}) discovery 
of the ribbon, 
could even have been predicted. Our explanation 
relies on the fact that the \emph{Voyager 1} and the undisturbed interstellar 
field directions (which we assume to be given by the \emph{IBEX} ribbon center (RC))  
share the same heliolatitude ($\sim$34.5\arcdeg) and are not 
far separated in longitude (difference $\sim$27\arcdeg). 
Our result confirms that \emph{Voyager 1} has indeed crossed the heliopause 
and offers the first independent confirmation that the \emph{IBEX} ribbon 
center is in fact the direction of the undisturbed interstellar magnetic 
field. For \emph{Voyager 2} we predict that the difference between the 
inner and the outer magnetic field directions at the heliopause will be 
significantly larger than the one observed by \emph{Voyager 1} 
 ($\sim$30\arcdeg \, instead of $\sim$20\arcdeg), and that the outer field
direction will be close to the RC.

\end{abstract}


\keywords{ISM: magnetic fields --- Sun: heliosphere}




\section{Introduction}

In August 2012, V1 crossed the HP and entered the local interstellar medium 
(LISM) \citep{burl13a,ston13}. Contrary to expectations, the change in the 
magnetic field direction across the HP was small \citep{burl13a} ($\sim$20\arcdeg). 
 As a result, the identification of the boundary as the HP remained
in doubt. 
This led to speculations about the existence of an unexpected transition 
layer at the HP that allows enhanced exchange of energetic particles with 
the interstellar medium while conserving magnetic field characteristics
typical for the heliosheath \citep{figl13,flor13,mcsc12,ston13,scmc13}. 
Only detection of electron plasma oscillations which allowed to evaluate 
the local plasma density at V1 in 2013, with values characteristic of the 
LISM, provided a strong evidence that the spacecraft has already crossed 
the heliopause \citep{gurn13}. However it is still not understood why the 
magnetic field direction observed beyond the HP is so close to the field 
measured in the inner heliosheath, and attempts are still being made to 
answer this question \citep{scmc13,figl13,flor13,opdr13,bopo14}.

In this letter we show how this observation can be explained in a simple 
and natural way if the undisturbed interstellar magnetic field (ISMF) far 
from the heliosphere points to the \emph{IBEX} RC \citep{funs09}.  
The ribbon is an almost circular arc in the sky of enhanced energetic neutral 
atoms emission discovered by the \emph{Interstellar Boundary Explorer} 
(\emph{IBEX}) \citep{mcco09,funs09}. The RC is widely interpreted as the 
direction of the undisturbed ISMF \citep{funs09,mcco13,funs13}.
 We use the RC location given by \cite{funs09} and \cite{mcco09} 
(see Table \ref{tab1}). The recent energy averaged RC position 
\citep{funs13} differs from it only by 1.6\arcdeg. A change by this 
amount would not affect our results significantly. 

There are two crucial ingredients of our explanation. First, the 
heliolatitudes of V1 and the undisturbed ISMF happen to be approximately 
the same ($\sim$34.5\arcdeg, see Table \ref{tab1}), with small 
angular distance between the 
two directions ($\sim$27\arcdeg). Second, for a strong interstellar field, 
the draping around the HP has a simple structure, the evolution of 
which (in the region of HP not far from the undisturbed field direction)
 can be traced for the field magnitude decreasing towards the realistic 
values. 

The essence of our argument is presented in Figure \ref{f1}.  The V1 and 
ISMF (RC) directions are shown on a sky map in orthogonal projection and 
Sun-centered equatorial (heliographic inertial) 
coordinates{\footnote{ http://omniweb.gsfc.nasa.gov/coho/helios/coor\_des.html}}. 
From V1 observations we know \citep{burl13a} that the magnetic field 
on the inner side of the HP is very close to the Parker spiral, that is to 
the -T direction in the RTN 
system{\footnote{  http://www.srl.caltech.edu/ACE/ASC/coordinate\_systems.html. 
The transformation from heliographic inertial to RTN coordinates 
follows from \cite{frha02}.}}.  
Projected on the sky map, the solar magnetic field line is tangent to 
the heliographic parallel passing through the V1 position. 
 B$_{\rm IN}$ and  B$_{\rm OUT}$ 
denote the directions of the magnetic field measured by V1 before and 
after the crossing the HP, respectively.

The projection of the ISMF line passing through V1 position must link
this point with the direction of the unperturbed ISMF (RC). If the draping
pattern is such that the projected ISMF line reaches RC from V1 with 
 little change in direction, the ISMF at V1 must be approximately parallel
to the inner field simply because V1 and RC have the same heliolatitude.

\section{Simple Draping Model}

\begin{figure} 
\epsscale{0.8}
\plotone{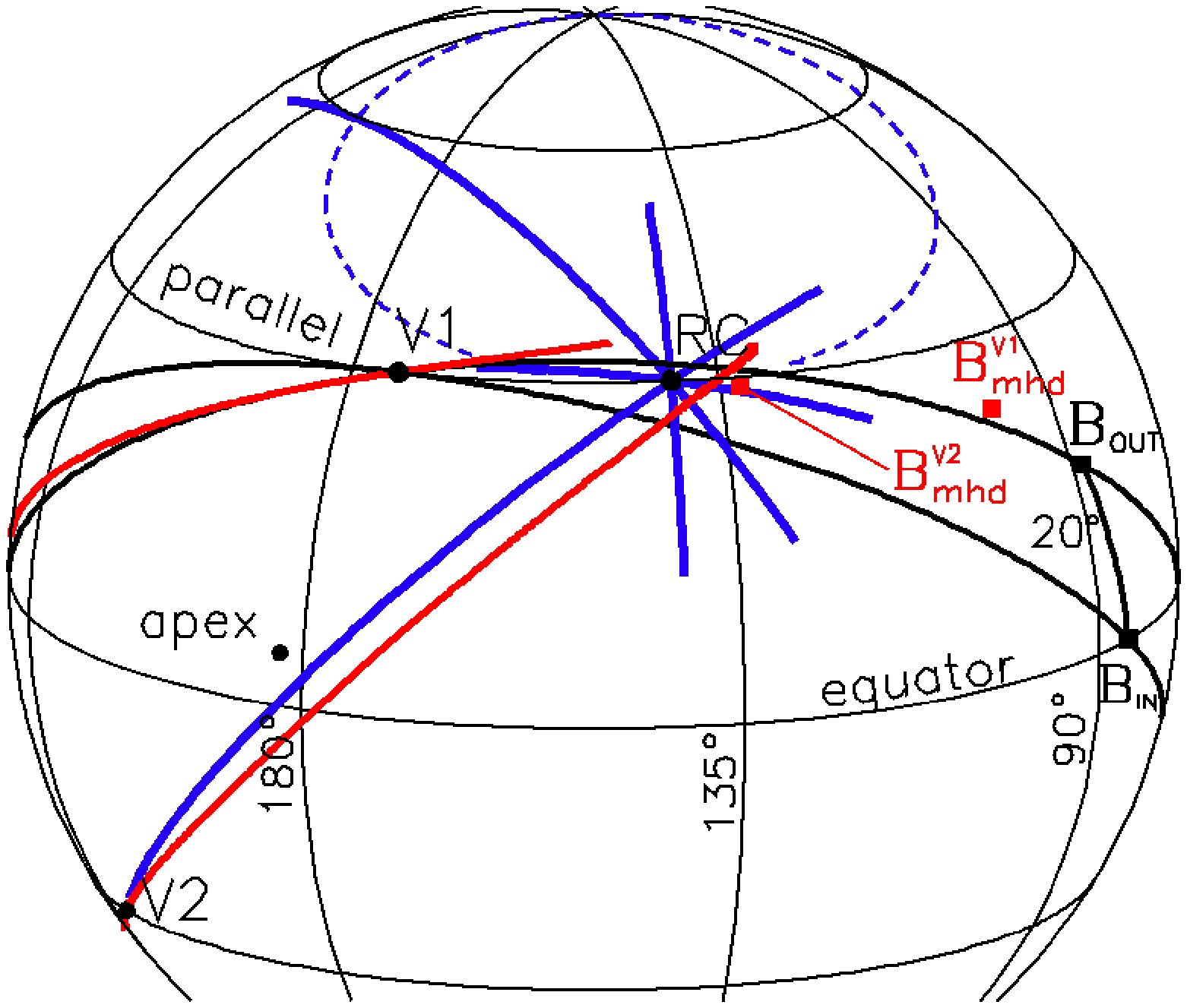}
\caption{Sky map in orthogonal projection and heliographic inertial coordinates. 
Shown are: \emph{Voyager 1} (V1), \emph{Voyager 2} (V2), \emph{IBEX} ribbon 
center (RC), interstellar inflow direction (apex), magnetic field directions 
measured by V1 before and after crossing the heliopause (B$_{IN}$ and B$_{OUT}$, 
respectively). Two great circles (black) link V1 with B$_{\rm OUT}$ and 
B$_{\rm IN}$ (the latter is tangent to the heliographic parallel at V1). 
Thick blue lines are the interstellar magnetic field lines 
in the idealized draping model. The dashed blue oval is a locus of points 
where the idealized draped interstellar field lines become tangent to the local 
heliographic parallel (illustrated by the extended field line in the north). 
The thick red lines are the draped magnetic field lines obtained from our 
simulation for 4$\mu$G case. The red squares (B$^{V1}_{mhd}$ and B$^{V2}_{mhd}$) 
are the directions of the interstellar field just outside the heliopause at  
directions of V1 and V2 for the same simulation.
\label{f1}}
\end{figure}

\begin{figure} 
\plotone{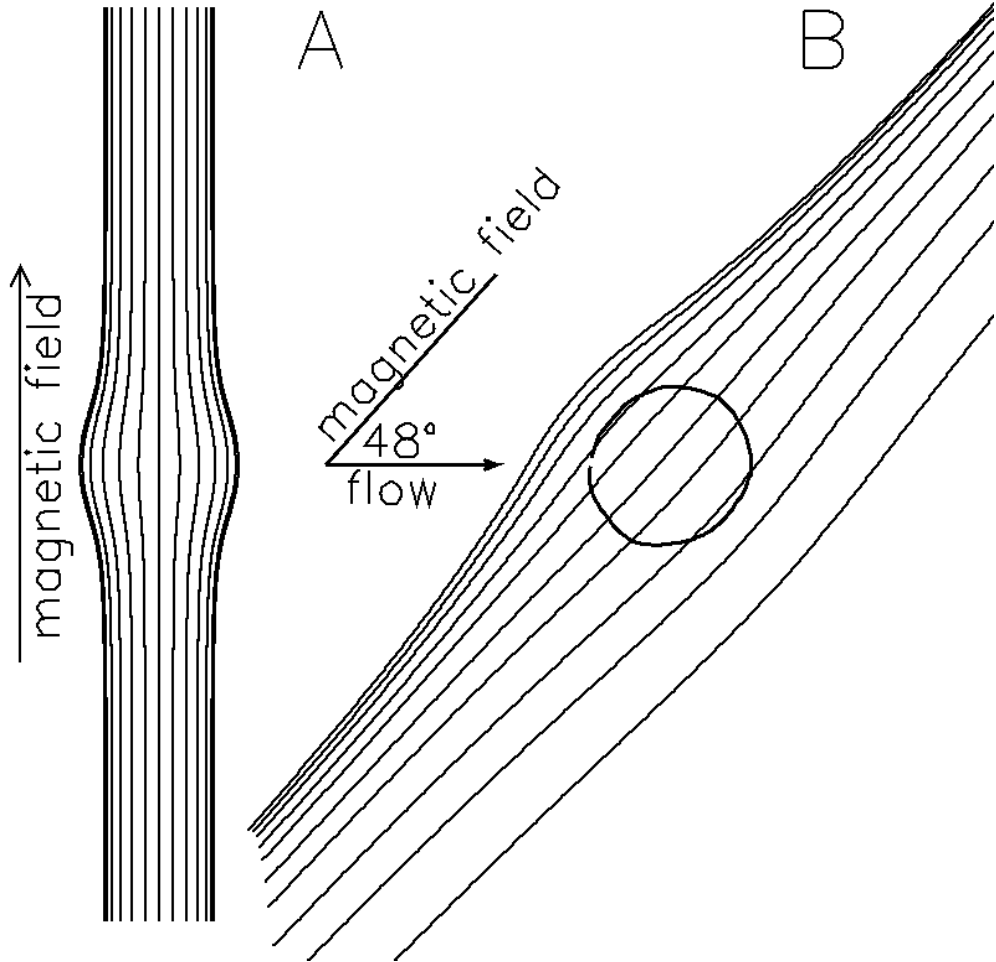}
\caption{Draped interstellar magnetic field lines: A) in Parker's solution 
(star at rest relative to the interstellar medium 
with insignificant plasma density), and B) obtained from 3DMHD simulations 
for spherically-symmetric solar wind and interstellar magnetic field 
of 20$\mu$G, inclined 48\arcdeg \,to interstellar plasma flow (projection on 
a plane which contains interstellar flow and field vectors). 
\label{f2}}
\end{figure}

The simplest ('idealized') draping model with this property is 
illustrated in Figure \ref{f1}, where the selected draped ISMF field lines are 
shown as thick blue lines. We chose one of them to pass through the V1 
and another through V2 position. In this model, based on the well-known 
solution by Parker \citep{park61}  (see also Figure \ref{f2}), the projected 
lines are arcs of great circles and originate in the undisturbed ISMF 
direction (RC).

In Parker's solution the star is at rest relative to the interstellar 
medium with insignificant plasma density and strong magnetic field.
 The stellar wind plasma flows away from the star, forming two streams, 
parallel and anti-parallel to the ISMF direction. 
Below we shall show, from our 3DMHD simulations \citep{czec13}, that 
the same idealized draping model can be regarded as the high ISMF 
strength limit for the case of the Sun, which is moving through the 
medium including the magnetized plasma. Moreover, when the ISMF 
strength is decreased to values actually observed by V1 beyond the 
HP (4-6 $\mu$G) \citep{burl13b}, or, 
since the draped field near the HP can be stronger than the
undisturbed ISMF (by $\sim{40}$\% 
if the field is as weak as $\sim$3$\mu$G), 
even to lower values (3 $\mu$G or 2.7 $\mu$G, 
\cite{gryg11, heer14, bjaf13}), 
the draped field near V1 direction 
in our heliographic projection remains similar enough in 
structure to the idealized model (Fig. \ref{f3}) to agree
approximately with V1 observations.  
The red line in Figure \ref{f1} shows 
the draped field line obtained for the ISMF of 4 $\mu$G. 
 The angle between the red line and the heliographic parallel 
passing through V1 direction
is larger than for the idealized draping model (the blue line). However,
this is consistent with V1 observations, which correspond to the small
but nonzero angle between the field outside and inside HP. 

To agree with the V1 measurements, the red line at V1 position should
be strictly tangent not to the heliographic parallel, but to the 
(V1, B$_{\rm OUT}$) plane (defined by the V1 direction and the direction
of the magnetic field measured by V1 outside the HP). In our projection
this plane is represented by the great circle passing through the points
V1 and B$_{\rm OUT}$. Figure \ref{f1} shows that the draped field line 
corresponding to our draping model for the ISMF strength of 4 $\mu$G
is indeed close to being tangent to this great circle. Moreover, the
draped field direction in space derived from the model B$^{V1}_{mhd}$ 
(the red square) is close ($\sim$10\arcdeg) to the measured field direction 
B$_{\rm OUT}$ (the black square).

Figure \ref{f1} also shows the heliographic parallel and the magnetic 
field line for the idealized draping model, passing through the V2 
position. The angle between the field line and the parallel is clearly 
large (30\arcdeg) implying that the solar magnetic field and the 
draped ISMF will not be close in direction along the V2 trajectory. 
Similar values ($\sim$ 32\arcdeg \,and 31\arcdeg) are obtained 
in our 3DMHD simulations 
for the ISMF strength of 4 and 3 $\mu$G, respectively. 
The angular distance between V2 position and the RC is also 
much larger ($\sim$97\arcdeg) than for V1.

 The vicinity of the V1 direction is not the only region where the
magnetic fields outside and inside the HP can be parallel (or antiparallel)
to each other. In the idealized draping model, each of the interstellar 
magnetic field 
lines become tangent to a heliographic parallel at two points. The 
dashed blue line in Figure \ref{f1} shows a locus of such points situated 
between the north pole and ISMF directions (a similar line appears in 
the southern hemisphere).  The draped field line which crosses this
oval becomes tangent to the heliographic parallel at the crossing point
(as illustrated by the elongated blue line in the north).
If a spacecraft trajectory direction would be 
close to this line, then, assuming the idealized draping model, 
the solar and the interstellar fields observed at 
the heliopause would be approximately parallel (or antiparallel, for 
opposite polarity of the solar field). The actual direction of Voyager 1 
trajectory (at the same solar latitude and close in longitude to the RC) 
satisfies this condition.

\section{MHD Models}

\begin{figure} 
\epsscale{0.9}
\plotone{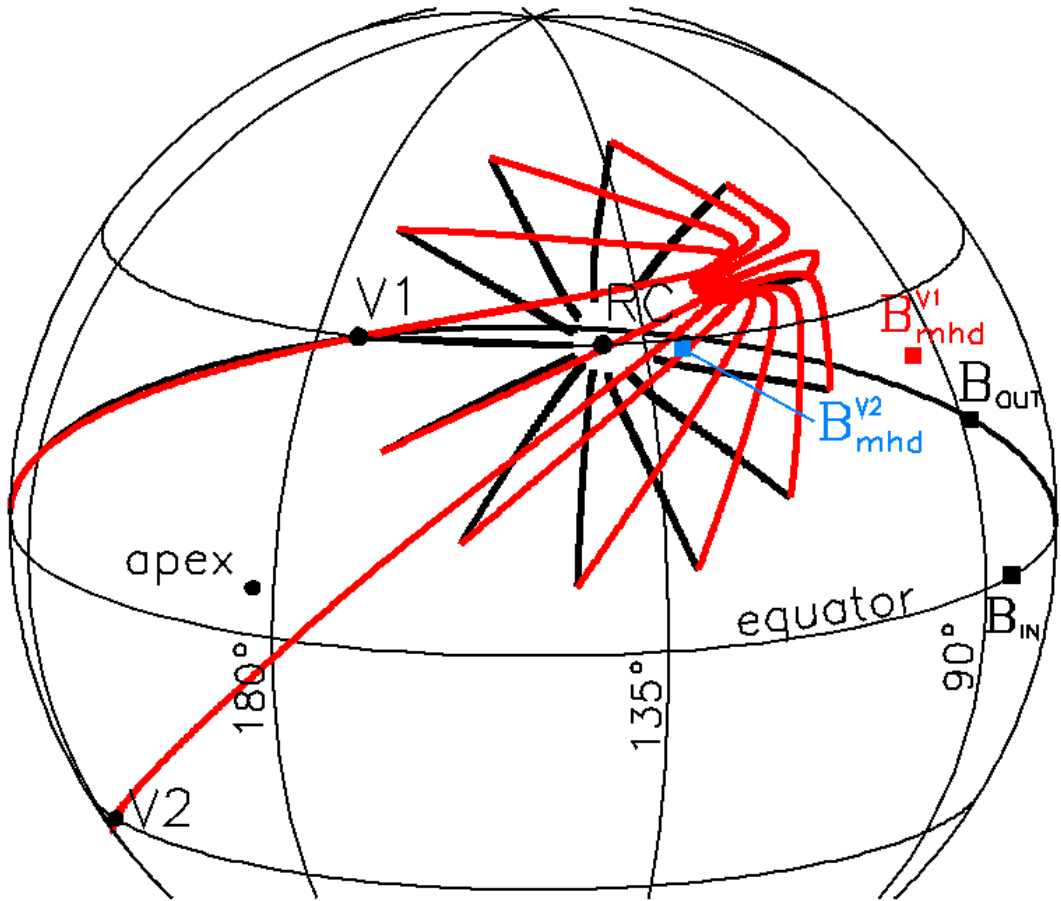}
\caption{Comparison of draped interstellar 
  magnetic field lines from our simulations for field strengths of 20 and 3 $\mu$G 
(black and red, respectively). The lines originate from selected points 
just outside the heliopause, distributed over a circle of radius 
27\arcdeg \,with the same center as the \emph{IBEX} ribbon.
\label{f3}}
\end{figure}

3DMHD simulations for the case of a star moving through a strongly 
magnetized (up to 20 $\mu$G) interstellar medium including also the 
plasma and neutral gas components \citep{czec13} led to a 
two-stream structure of the stellar plasma flow. In this case 
the streams (containing most of the plasma flow) are slightly deflected 
from the interstellar field direction. The rest of the outflowing 
stellar plasma forms a tail opposite to the stellar motion.
 The two stream structure becomes less pronounced for larger
values of the neutral interstellar hydrogen density. 
 Nevertheless, the draping structure for the strong interstellar 
field case (Fig. \ref{f3}, black arcs) remains similar to 
 our simple draping model (Fig. \ref{f1}, blue lines).

After the crossing of the HP, V1 is observing the interstellar field 
strength between 4 and 6 $\mu$G. To check the draping structure for the 
ISMF in this range we performed 3DMHD simulation for the local field 
strengths of 6, 5, 4, 3, and 2.7 $\mu$G, with the asymptotic field 
directed towards the \emph{IBEX} RC. 
We found that the weaker the field the more its draping structure 
differs from the one of the strong (20 $\mu$G) field case. 
Figure \ref{f3} illustrates this difference for the ISMF strength 
of 3 $\mu$G. Although the draped field is not as simple as for 
the 20 $\mu$G, in our region of interest (between the undisturbed 
field (RC) and V1 direction) its structure simplifies and takes 
the form consistent with V1 magnetic field measurements. 
As explained above, the draped magnetic field line passing through 
the V1 trajectory is close to tangent to the plane defined by 
the V1 and B$_{\rm OUT}$ directions, and  close in direction to the 
solar magnetic field on the inner side of the HP.

  The direction of the draped magnetic field in space at the V1 
position following from our models depends on the assumed ISMF strength.
For a very strong field (20 $\mu$G) it is different from the V1 result 
B$_{\rm OUT}$, but for the weaker and more realistic ISMF strength 
values it becomes close to B$_{\rm OUT}$ (within about 10\arcdeg). 
At the same time, the angle between the draped magnetic field at V1 
and the heliographic parallel increases from a low value (less than 
10\arcdeg) obtained for the idealized draping model and for strong 
ISMF (20 $\mu$G) to a somewhat higher value (15\arcdeg, close to 
V1 observations) for the realistic ISMF.

In our calculations we use 3DMHD time-stationary numerical code
describing the interaction of the solar wind with the LISM  
(boundary conditions for our models are shown in Table \ref{tab2}). 
The parameters of the solar wind are based on 
Ulysses data \citep{eber09}. The interstellar medium parameters were 
chosen to agree with observations from the \emph{IBEX} and \emph{Voyager 1} 
spacecraft \citep{gurn13}, and model estimations. 
 The code employs a fixed grid.
A uniform flux of neutral hydrogen is assumed \citep{bzow09,ragr08}. 
The solar magnetic field is disregarded to limit the undesired numerical 
effects (like artificial diffusion of the magnetic field at the 
HP) that could affect the heliopause region. 
 In effect we assume that, as in ideal MHD case, there is no 
coupling between the magnetic fields across the HP.  

 We use the models to obtain the shape of the HP and
the structure of the draped magnetic field. Since the HP
in our models is not a discontinuity but a transition of finite 
width, we cannot reliably calculate the behaviour of quantities 
like the magnetic field strength or plasma density with
distance in vicinity of the HP.    
For the magnetic field direction
we find that the 
agreement of the 4 $\mu$G and 3 $\mu$G cases with V1 observations 
is satisfactory. The calculated ISMF direction at V1 just outside the HP 
(the red squares marked  B$^{V1}_{\rm mhd}$ in Figure \ref{f1} 
and Figure \ref{f3}) are
deflected by $\sim$ 11$^{o}$ 
from the observed direction.
While the distance to the HP is overestimated, the obtained termination shock 
distances at V1 and V2 are within few percent 
from the measured ones
\citep{ston05,ston08}
when the slow wind speed is taken to be 500 km/s, but smaller (by less than
20 \%
for the slow wind speed of 400 km/s.  

\section{Discussion and Conclusions}

Throughout this paper we are using the heliographic projection.
We are interested in the magnetic field lines near the heliopause. 
The inner and the outer magnetic fields are then draped over the same 
surface. Since the heliopause is not strictly perpendicular to the 
radial direction, the angles between the projected field lines are not 
the same as the angles between the field lines in space. However, as 
long as we exclude the case of a heliopause tangent to the radial 
direction, the lines draped over the same element of the heliopause and
 strictly parallel (antiparallel) in projection have the same property 
also in space. We can therefore restrict attention to the heliographic 
projection and do not have to specify the shape of the actual heliopause. 
If the considered part of the heliopause is not very far from being 
perpendicular to radial direction, the field lines forming a small 
angle in projection, will also form a small angle in space.

Our results clarify the cause of similar magnetic field 
directions on both sides of the HP, as observed by V1: the heliographic 
latitude of V1 happens to be the same as that of the undisturbed ISMF 
(to be given by the \emph{IBEX} RC). We have shown that for the 
 wide range of the ISMF strength of values, including those 
actually observed by V1 beyond the HP, the draped interstellar field 
near V1 trajectory (calculated with zero heliospheric magnetic 
field to avoid artificial field diffusion across the boundary)
must then be close in direction to the solar inner heliosheath magnetic 
field. For realistic ISMF strength values, our model results 
obtained for the draped field direction in space are within 10\arcdeg  
\,from V1 observations beyond the HP. 
For \emph{Voyager 2} we predict that the difference between the inner 
and the outer magnetic field directions at the heliopause will be 
significantly larger than the one observed by \emph{Voyager 1}
 ($\sim$30\arcdeg \,instead of $\sim$20\arcdeg) and that the outer 
field direction will be close (7\arcdeg \,- 9\arcdeg \,off) to the RC. 

The explanation is natural and simple, and removes the only objection 
against the claim that V1 reached the interstellar medium. It supports 
the interpretation of the ribbon center as the direction of the 
undisturbed ISMF, with the ISMF oriented towards the northern hemisphere
It is consistent with the classical picture of the 
HP not dominated by the turbulence or reconnection effects.
The reasonable agreement between V1 and our results for the
magnetic field direction gives us confidence that the results are not 
affected by the approximations inherent to our models. 

 Approximate agreement with V1 magnetic field measurements was also
achieved in numerical models that differ significantly from our approach.
\cite{bopo14} (see also \cite{pogo09}), who take
HP instability into account, obtain a solution in which V1 is moving
through a local interstellar plasma intrusion, partly surrounded by
the heliosheath plasma. They predict that V1 may re-enter the heliosheath
in the near future. \cite{opdr13}, from numerical calculations that include
both solar and interstellar magnetic fields, conclude that the presence of
solar magnetic field strongly affects the draping of the ISMF over the
HP. In their solution, for different asumptions about the undisturbed ISMF,
the draped ISMF is close in direction to the heliospheric
magnetic field in a large area of the HP including also the V1 and V2 trajectory
directions. Calculations of \cite{bopo14} do not confirm this
result. The solar field assumed by \cite{opdr13} is very simplified.
In particular, the sector polarity structure is ignored.
Contrary to \cite{opdr13}, we assume that the solar magnetic field does 
not strongly
affect the shape of the HP and the draping pattern. The similarity of the
field directions on both sides of the HP follows from the relative directions 
of V1 and the undisturbed ISMF.





\acknowledgments

This work was supported by the Polish National Science Center grant 2012-06-M-ST9-00455.

\begin{table}
\begin{center}
\caption{Characteristic directions in galactic (GAL), solar ecliptic (SE), 
 solar equatorial (HGI\tablenotemark{a}), and RTN\tablenotemark{b} coordinates (J2000).\label{tab1}}
\begin{tabular}{ l | c c | c c | c c | c c } 
\tableline\tableline          
& \multicolumn{2}{c|}{GAL} & \multicolumn{2}{c|}{ SE} & \multicolumn{2}{c|}{HGI\tablenotemark{a}} & \multicolumn{2}{c}{RTN\tablenotemark{b}}\\ 
\tableline
 &  lon  &  lat  &  lon  &  lat  &  lon  &  lat  &  lon  &  lat \\
\tableline
 & (deg) & (deg) & (deg) & (deg) & (deg) & (deg) & (deg) & (deg) \\
\tableline                       
$\rm V1$           &  32.7 &  28.1 & 254.9 & 35.0 & 174.1 & 34.5  &  & \\ 
 $\rm V2$          & 342.1 & -31.2 & 289.4 & -34.2 & 217.2 & -30.0   &  & \\
$\rm RC (=B_{\rm IS})$ & 32.9 & 55.3 & 221.0 & 39.0 & 140.9 & 34.6  &  &  \\   
$\rm apex$         &  5.3 & 12.0 & 259.0 & 4.98 & 182.6 & 5.4   &  & \\   
$\rm B_{\rm IN}$       & 232.0 & 60.4 & 159.6 &  7.2 & 83.9 &  0.0  & 270. &  0. \\   
$\rm B_{\rm OUT}$      & 197.2 & 75.1 & 163.7 & 26.0 & 88.0 & 18.8  & 284. & 13. \\    
\tableline
$B^{\rm V1}_{\rm MHD} \ \ 2.7 \mu$G\tablenotemark{1} & 155. & 80. & 169. & 35. & 93.  & 28. & 293. & 18. \\   
$B^{\rm V1}_{\rm MHD} \ \ 3 \mu$G\tablenotemark{1}& 164. & 73. & 161. & 35. &  95.  & 28.  & 295. & 17. \\   
$B^{\rm V1}_{\rm MHD} \ \ 4 \mu$G\tablenotemark{1} & 168. & 53. & 137. & 34. & 100.  & 27. & 298. & 13.  \\   
$B^{\rm V1}_{\rm MHD} \ \ 20.0 \mu$G & 28. & 63. & 211.  & 36. & 132.  & 31.  & 325. &  3. \\   
$B^{\rm V1}_{\rm MHD} \ \ 3 \mu$G\tablenotemark{2} & 159. & 71. & 159. & 37. &  93.  & 30. & 294. & 19.  \\   
$B^{\rm V1}_{\rm MHD} \ \ 4 \mu$G\tablenotemark{2} & 126. & 83. & 176. & 36. &  99.  & 29. & 299. & 16.  \\   
\tableline
$\rm B^{\rm V2}_{\rm MHD} \ \ 2.7 \mu$G\tablenotemark{1} & 36. & 65. & 208. & 39. & 129.  & 34.  & 253. & 29. \\   
$\rm B^{\rm V2}_{\rm MHD} \ \ 3 \mu$G\tablenotemark{1} & 35. & 64. & 209. & 39. & 130.  & 33.  & 254. &  30. \\   
$\rm B^{\rm V2}_{\rm MHD} \ \ 4 \mu$G\tablenotemark{1} & 34. & 62. & 212. & 39. & 132.  & 34.  & 256. & 31. \\   
$B^{\rm V2}_{\rm MHD} \ \ 20.0 \mu$G & 30. & 53. & 224. & 37. & 143.  & 33. & 265. & 36. \\   
$\rm B^{\rm V2}_{\rm MHD} \ \ 3 \mu$G\tablenotemark{2} & 37. & 62. & 212. & 40. & 133.  & 35. & 255.  & 32.  \\   
$\rm B^{\rm V2}_{\rm MHD} \ \ 4 \mu$G\tablenotemark{2} & 35. & 60. & 215. & 40. & 135.  & 35.  & 257. & 33. \\   
\tableline                                  
\end{tabular}
\begin{list}{}{}
\item[$^a$]http://omniweb.gsfc.nasa.gov/coho/helios/coor\_des.html
\item[$^b$]http://www.srl.caltech.edu/ACE/ASC/coordinate\_systems.html
\item[$^ 1$] Table \ref{tab2} solar wind case 400/750.
\item[$^ 2$] Table \ref{tab2} solar wind case 500/750.
\end{list}
\end{center}
\end{table}
\
\begin{table}
\begin{center}
\caption{Boundary conditions for MHD calculations\tablenotemark{1}\label{tab2}}         
\begin{tabular}{c c c c c | c c} 
\tableline\tableline          
\multicolumn{5}{c|}{LISM parameters} & \multicolumn{2}{c}{Solar Wind parameters\tablenotemark{2}}\\
\tableline
$\rm B_{\rm IS}$ & $\rm T_{IS}$ & $\rm n_{IS}$ & $\rm n_H$ & $\rm V_{IS}$ & $\rm V_{SW, 1 AU}$ & $\rm n_{SW, 1 AU}$ \\ 
\tableline
($\rm \mu$G) & (K) & (cm$^{\rm -3}$) & (cm$^{\rm -3}$) & (km/s) & (km/s) & (cm$^{\rm -3}$) \\ 
& & & & & slow/fast & slow/fast \\ 
\tableline                        
   20    & 6300 & 0.04 & 0.1 & 23.2 & 750 & 4.2\\   
  4, 5, 6 & 6300 &  0.06 & 0.1 & 23.2 & 400 & 5.55\\
  2.7, 3, 4, 5 & 6300 &  0.06 & 0.1 & 23.2 & 400/750 & 5.55/1.58\\ 
  3, 4 & 6300 &  0.06 & 0.1 & 23.2 & 500/750 & 5.55/2.46\\
\tableline                                   
\end{tabular}
\begin{list}{}{}
\item[$^1$]The inner boundary of the calculation region is at 15 AU, the outer at 5849 AU from the Sun.  
\item[$^2$]In the slow/fast solar wind cases, the slow wind flows within 36\arcdeg \,from the solar equator.
\end{list}
\end{center}
\end{table}

\end{document}